\documentclass[conference]{IEEEtran}
\IEEEoverridecommandlockouts

\def\BibTeX{{\rm B\kern-.05em{\sc i\kern-.025em b}\kern-.08em
    T\kern-.1667em\lower.7ex\hbox{E}\kern-.125emX}}

\usepackage{amsmath,graphicx}
\usepackage{amssymb}

\usepackage{comment}

\graphicspath{{FIG/}}
\usepackage{subfigure}
\usepackage{color}

\title{Harmonic MUSIC Method for mmWave Radar-based Vital Sign Estimation}

\author{\IEEEauthorblockN{Chieh-Hsun Hsieh, Tung-Lin Tsai, and Po-Hsuan Tseng}
\IEEEauthorblockA{Department of Electronic Engineering, \\
National Taipei University of Technology,\\Taipei, Taiwan}
}

\begin{document}
%
\maketitle
\begin{abstract}

This paper investigates the application of millimeter-wave (mmWave) radar for the estimation of human vital signs. Aiming to obtain more accurate frequency estimation for periodic signals of respiration and heartbeat, we propose the harmonic MUSIC (HMUSIC) algorithm to consider harmonic components for frequency estimation of vital sign signals. 
In the experiments, we tested different subjects' vital signs. 
Experimental results demonstrate that the 89-th percentile errors in respiration rate and the 88-th percentile errors in heartbeat rate are less than 3 respirations per minute and 5 beats per minute.

\end{abstract}
\begin{IEEEkeywords}
millimeter-wave radar, vital sign estimation, static clutter removal, MUSIC
\end{IEEEkeywords}

\section{Introduction}
\label{sec:intro}

In the landscape of vehicular technology, advanced driver assistance systems (ADAS) are indispensable for ensuring driver safety and adding convenience. These systems are engineered to swiftly intervene during emergencies by reliably tracking critical vital signs such as the driver's respiration and heart rates. Unlike wearable devices, often unsuitable for prolonged monitoring, non-contact radar-based techniques offer a wireless solution for accurately estimating vital signs. Millimeter-wave frequency-modulated continuous wave (FMCW) radar 
\cite{gharamohammadi2023vehicle,hsieh2023mmwave,yu2024mmwave} stands out among the available options. It excels in detecting the subtle chest wall movements caused by breathing and heartbeat within the vehicle, and it also measures the distance and velocity of other vehicles. Consequently, FMCW radar has become a cornerstone technology in ADAS applications.

Under FMCW radar, the correct range-azimuth bin containing vital signals ({\it vital bin}) 
must first be estimated before vital sign estimation. Different human target localization assumptions and scenarios have been considered. Some wireless vital signs monitoring is designed to track the breath and heart rate through a known distance position, which could be used to monitor the driver. Besides, the current literature on vital sign estimation based on FMCW radar mainly focuses on single-person scenarios \cite{ref4_MPS, ref5_MPC, ref11_PiViMO}, or on vital sign estimation and target localization in relatively open spaces \cite{ref1_SOD, ref3_mmHRV, ref5_MPC, ref8_TI, ref9_rangeint}, without any clutters. 
In this paper, the position of humans is assumed to be unknown, where human target localization is required in cluttered environments \cite{hsieh2022mmwave} for the correct range-azimuth bin detection and its phase compensation before vital sign estimation.

After the phase compensation, the accuracy faces limitations attributed to harmonics from non-sinusoidal periodic signals by their presentation in respiration and heartbeat frequencies. These non-sinusoidal periodic signals contain components that can interfere with heartbeat estimation \cite{multi_pitch, NLS_vs}. 
Due to the harmonics or inter-modulation between breathing and heartbeat components \cite{ref4_MPS, ref8_TI, ref10_MRC}, 
conventional techniques like fast Fourier transform (FFT) are challenged. To address multi-frequency signal decomposition approaches like variational mode decomposition (VMD) \cite{ref3_mmHRV} have been used to decompose narrowband signals into intrinsic mode functions. However, vital sign signals often contain multiple harmonic frequencies, requiring further selection for the desired heartbeat frequency component. A nonlinear least squares (NLS) framework with harmonic consideration \cite{NLS_vs} has been applied to harmonic power spectral density for heartbeat and respiration rate estimation.

In this paper, we estimate vital signs in a cluttered environment with an unknown human target position. Our research centers on developing the harmonic multiple signal classification (MUSIC) method, which leverages the harmonics in two vital-sign sources, i.e., respiration and heartbeat frequencies. In contrast, earlier works such as NLS \cite{NLS_vs} focus exclusively on applying NLS formulation with harmonic consideration for vital sign estimation, while our work applies the super-resolution-based algorithm \cite{schmidt1986multiple}  based on signal correlation. We propose a novel HMUSIC method that was not explored in these earlier investigations of vital sign estimation. 
In the experimental results, HMUSIC outperforms several state-of-the-art methods, such as MPC \cite{ref5_MPC}, NLS \cite{NLS_vs}, regarding estimation accuracy.

\section{Signal Model}

\subsection{FMCW Radar Signal}
Assuming a complex-valued FMCW radar transmission antenna (Tx) emits a chirp signal
\begin{align}
    s_{T}(t)= A_{T}  \exp\{j (2 \pi f_{0} t + \pi S t^2) \}, \ 0 < t < T_c \ ,
\end{align}
where $f_0$ represents the initial frequency, $A_{T}$ is the amplitude of the transmitted signal, $S$ denotes the modulation frequency slope, and $T_c$ is the end time of a single chirp. 
To characterize between static clutter and the radar's distance from human targets in the model, the distance between the $k$-th target and the radar is denoted as $R_k$, while the distance change over time, represented as $x_k(t)$, characterizes the respiratory and heartbeat-related motion of a human target. Therefore, the distance from the target to the radar can be expressed as a function of time as 
\begin{align}
    R(t) =  R_0 + x_0(t), & &  \text{for human target}.
\end{align}
As the transmitted signal traverses a time delay $t_0 = 2R(t)/c$, it represents the round-trip time for the target to bounce the signal back to the receiving antenna (Rx), where $c$ is the speed of light. The radar's received signal is given by $A s_T(t-t_0)$, where $A$ is the amplitude of the received signal. FMCW radar utilizes the time delay property by mixing the transmitted signal with the complex conjugate of the received signal. This process generates the intermediate frequency signal, which is then subjected to low-pass filtering to extract the frequency difference signal:
\begin{align}
    s_{IF}(t) &= s(t)  [A_k s_T^{*}(t-t_0)], \ \ t_0 < t < T_c  \nonumber \\
    & \approx  A_0 A_T \exp\{ j(2\pi f_0 t_0 + 2 \pi S t_0 t)\} \nonumber \\
     & \approx  A_0 A_T \exp\{j [\omega_{b} t +  \frac {4 \pi} {\lambda_0} R(t) ] \}  \ ,
\end{align}
$\omega_{b} = 4\pi \frac {S R_0} {c}$ and remaining $\frac {4 \pi} {\lambda_0} R(t)$ represent the beat frequency and phase of the mixed signal from the $k$-th target, with the assumption of a total of $K$ targets. 
$\lambda_0$ is the wavelength of the FMCW radar's starting frequency. 
Notice that the second approximate equality omits $\pi S t_0^2$, and the third term ignores the $x_0(t) \cdot t$ term as discussed in \cite{arctan_dem}. 
Additionally, for a stationary human target, the distance variation caused by breathing and heartbeat within the chest can be considered unchanged during a single chirp duration $T_c$. Thus, the beat frequency term is represented by the distance of the target $R_0$. 
Frequency estimation of this mixed signal can be utilized to compute the distance between the target and the radar, given by $R_0 = \frac {\omega_{b} c} {4 \pi S}$.


\subsection{Digital Signal Model} 
To facilitate digital signal processing, analog-to-digital conversion (ADC) is performed within each chirp with a fast-time sampling interval of $T_f$, resulting in a total of $N_f$ samples. Due to the frequency of respiration and heartbeat, it is not sufficient to observe the phase changes caused by human vital signs within a single chirp time. 
Apart from fast-time sampling, slow-time sampling is conducted among different frames with an interval of $T_s$ to capture the phase changes in a total of $N_s$ samples, thereby extracting human vital sign information. The ADC sampling at the $n$-th point in fast time and the $m$-th frame in slow time are then reformulated as 
\begin{align}
    y_m[n] = \tilde A_0 \exp\{ j [n \omega_{b} T_{f} + \frac {4 \pi} {\lambda_0} R(m T_s) ] \} 
    \label{eq:m_th_ADC} 
\end{align}
$n = 0, 1, ..., N_f - 1$ and $m = 0, 1, ..., N_s - 1$ represent the index values of fast-time samples and slow-time samples. 
$\tilde A$ denotes the amplitude associated with transmitted and received signals. While signal amplitude can be affected by factors like path loss and radar cross-section changes due to moving objects at varying distances, these effects are not evident compared to phase information. Therefore, amplitude information is not considered for estimating human vital signs. Besides, since we consider the uniform linear antenna array for the MIMO radar, we denote the ADC sample of the $v$-th antenna based on the definition in \eqref{eq:m_th_ADC} as 
\begin{align}
    y_m[v,n]= \tilde{A}_{v} y_m[n] \cdot \exp\{ j v \frac{2 \pi d}{\lambda_0} \sin \theta  \} 
    \label{eq:v_th_ADC}
\end{align}
where $v = 0, 1, ..., N_v-1$ is the index of antennas, and $\theta$ is the azimuth angle of arrival. 
$d$ is the antenna spacing with the configuration of $d = \lambda_0 / 2$. 
The additional phase term caused by the arrival angle is summarized as $v \pi \sin \theta$ compared to other antenna. 
In the slow-time sampling context, target objects undergo phase estimation after distance estimation. It's possible to distinguish between static objects and human targets by assessing phase changes, given that the vital sign signal $x_0(m T_s)$ is embedded within the phase.

\subsection{Vital Sign Signal Model}
To simplify the signal model for real-time computations, various studies have approximated vital sign signals as pure sinusoids or periodic signal models containing multiple harmonic components \cite{NLS_vs}. 
Moreover, literature has explored modeling the distance variations caused by respiration and heartbeat in the chest cavity 
\cite{ref11_PiViMO}, 
yet these types of mathematical models often encounter challenges in accurately estimating parameters during real-time radar signal processing. 

In this study, the human vital sign signal is approximated as a periodic signal composed of $L$ harmonics as 
\begin{align}
    x(t) &= \sum_{l=1}^{L} M_{r_{l}} \cos( l \cdot\omega_r t) + \sum_{l=1}^{L} M_{h_{l}} \cos( l \cdot \omega_h t) , \label{eq:vs_signal}
\end{align}
where $f_r= \omega_r/ (2\pi)$ and $f_h = \omega_h / (2\pi)$  are the fundamental frequencies of the respiration and heartbeat, and $M_{r_{l}}$ and $M_{h_{l}}$ are their amplitudes.

\section{Proposed Harmonic MUSIC (HMUSIC)}
\begin{figure}
    \centering
    \includegraphics[width=3.3in]{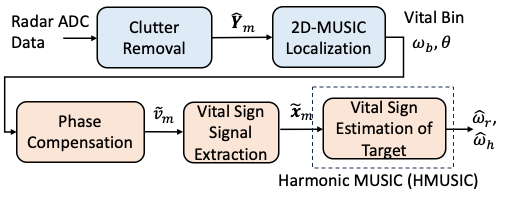}
    \caption{Flowchart of Proposed HMUSIC Algorithm}
    \label{fig:flowchart}
\end{figure}
Fig. \ref{fig:flowchart} illustrates the proposed system diagram, named after the vital sign estimation part -- HMUSIC algorithm. The radar-received ADC signal is first processed with the clutter removal algorithm. Then, the 2D MUSIC localization is employed to estimate the human target's range and angle. By compensating for the phase caused by the human's position, the remaining phase from the targeted range-angle bin is then extracted according to the differentiate and cross-multiply
(DACM) \cite{DACM} to avoid phase discontinuities. The designed HMUSIC considered respiration and the heartbeat and their harmonic terms in model assumption and classified these two sources to obtain a better estimation. 

\subsection{Static Clutter Removal}
To distinguish static clutters from the human target, static clutter removal is performed using the background clutter estimation through a moving averaging filter \cite{FMCW_MTI} for the ADC raw data. This process results in a signal $\hat{\mathbf{Y}}_m$ that contains only the human targets, who breathe regularly, causing chest movement, and would not be regarded as static clutter. 


\subsection{2D-MUSIC-based Localization}
To facilitate target localization and consider the target as static within a frame, 
\eqref{eq:v_th_ADC} is represented in matrix form as $\hat{\mathbf{Y}}_m \in \mathbb {C}^{N_v\times N_f}$ for subsequent matrix operations.
\begin{align} 
    \hat{\mathbf{Y}}_m &= \Big[ \hat{y}_m[v,n] \Big] = \tilde A_m e^{j\psi_{m}} \mathbf{a}(\theta) \mathbf{s}^T(\omega_{b})+ \mathbf W_m  \ , \label{eq:data_matrix} \\
    \mathbf{a}(\theta)&=\left[1,\ e^{j \pi \sin\theta}, \dots, \ e^{j \pi (N_v - 1)\sin\theta} \right]^{T}  \in \mathbb 
    {C}^{N_v\times 1} \ ,
    \\ 
    \mathbf{s}(\omega_{b})&= \left[1,e^{j\omega_{b}T_f},..., e^{j\omega_{b}(N_f - 1)T_f} \right] ^{T}  \in \mathbb {C}^{N_f \times 1} \ .
\end{align}
$\mathbf{a}(\theta)$ is the angle steering vector formed by $N_v$ antennas. $\mathbf s(\omega_{b})$ represents the beat frequency-related (distance-related) steering vector formed by $N_f$ fast-time samples. 
$\mathbf W_m$ is the $N_v \times N_f$ additive Gaussian white noise combined with potentially other static clutter. Besides the phase caused by the position of the human target, $\psi_{m}$ is the remaining phase term related to the vital sign in this estimation problem. 

Then, we employ the 2D-MUSIC algorithm \cite{2DMUSIC_FFT}, which distinguishes signal components from noise in the autocorrelation matrix of the signal. It then performs pseudo-spectrum estimation on these vector spaces to obtain high-resolution localization information. For the matrix $\hat{\mathbf{Y}}_m$, it is transformed into a space-time vector $\tilde {\mathbf Y}_m$ across antennas and fast time samples as
\begin{align}
    \tilde {\mathbf Y}_m &= \Big[\hat{\mathbf{y}}_m[0] , \hat{\mathbf{y}}_m[1], ..., \hat{\mathbf{y}}_m[N_v-1]\Big]^T \in \mathbb C^{N_v N_f \times 1} \ , 
\end{align}
where $\hat{\mathbf{y}}_m[v] = \Big[y_m[v,0],\ y_m[v,1],\ ...,\ y_m[v,N_f-1]\Big]$. 
The autocorrelation matrix of this space-time vector, along with its eigenvalues sorted in descending order, can be represented as 
\begin{align}
    \mathbf R_{\tilde {\mathbf Y}_m} = \mathbb E \{\tilde {\mathbf Y}_m \tilde {\mathbf Y}_m^H \}  = \mathbf U_{\tilde {\mathbf Y}_m} \pmb \Lambda_{\tilde {\mathbf Y}_m} \mathbf U_{\tilde {\mathbf Y}_m}^H \ ,
\end{align}
with its diagonal matrix $\pmb \Lambda_{\tilde {\mathbf Y}m}$ and corresponding eigenvector matrix $\mathbf U_{\tilde {\mathbf Y}_m}$, whose $i$-th column is the eigenvector $\mathbf{g}_i$ of $\mathbf R_{\tilde {\mathbf Y}_m}$. 
By considering one human target, the noise subspace consisting of $N_f N_v - 1$ feature vectors 
is represented as $\mathbf G_{\tilde {\mathbf Y}m} = \left[ \mathbf{g}_2,\mathbf{g}_3, ..., \mathbf{g}_{N_v N_f} \right] \in \mathbb C^{(N_v N_f) \times (N_f N_v - 1)}$. This subspace is subsequently used to evaluate the 2D MUSIC pseudo-spectrum for range and angle estimation:
\begin{align}
\{\hat{\theta}, \hat{\omega_b}\} = \arg \max_{\theta, \omega_b}   \frac {1} {\|\mathbf V^H(\theta, \omega_b) \mathbf G_{\tilde {\mathbf Y}_m}\|_{F}^2} \ .\label{eq:Loc_MUSIC}
\end{align}
$|| \cdot ||_{\text F}$ represents the Frobenius norm, and the corresponding space-time vector $\mathbf V(\theta, \omega_b) = \mathbf a(\theta) \otimes \mathbf s(\omega_b)$, where $\otimes$ denotes the Kronecker product. 
The estimated $\hat{\omega_b}$ and $\hat{\theta}$ composite the range-azimuth bin (vital bin) of the human.

\subsection{Phase Compensation using Estimated Range and Angle}
After obtaining the location of the human target, i.e., the vital bin, we further estimate the remaining phase term $\psi_{m}$ related to the vital sign in this bin. In \eqref{eq:data_matrix}, we first define ${\mathbf u}_m = \tilde A_m e^{j\psi_{m}} \mathbf{a}(\theta)$ 
 and estimate it from the measurement $\mathbf Y_m$ using the least squares criteria: 
 $\hat{\mathbf{u}}_m =\arg \min_{{\mathbf u}_m} ||\mathbf Y_m^T - \mathbf{s}(\omega_{b}) {\mathbf u}_m^T ||_{F}^2$. 
This leads to $\hat{\mathbf{u}}_m^T = [\mathbf{s}^H(\omega_{b}) \mathbf{s}(\omega_{b})]^{-1} \mathbf{s}^H(\omega_{b}) \mathbf Y_m^T$, where $\mathbf \mathbf{s}^H(\omega_{b}) \mathbf \mathbf{s}(\omega_{b}) = N_f$. Therefore, the amplitude and phase of the $m$-th point in slow time $\hat{{\mathbf u}}_m$ is represented as 
\begin{align}
    \hat{{\mathbf u}}_m = \frac {1} {N_f} \mathbf Y_m \mathbf s(\omega_{b}) \ . \label{eq:chs_IQ}
\end{align}
Subsequently, the relevant amplitude and phase of the original signal are estimated while excluding the angle of the target using the least squares method:
\begin{align}
    \tilde{{ v}}_m = \frac {1} {N_v} \mathbf a^H(\theta) \hat{{\mathbf u}}_m \ .
\end{align}
Although arctangent demodulation \cite{arctan_dem} can be used to obtain the phase information of human vital sign signals from $\tilde{{ v}}_m$, the phase obtained from the arctangent operation exhibits discontinuities at the boundaries of $\pi$ to $-\pi$. We first employ the maximum likelihood estimation \cite{remote_monitor} to eliminate DC offsets in $\tilde{{v}}_m$, ensuring precise phase information.
Besides, the phase discontinuities can be overcome through techniques like phase unwrapping or DACM \cite{DACM}. 
The vital-sign relevant phase signal $\phi_m$ in \eqref{eq:data_matrix} is estimated by DCAM as ${\tilde x}_m$.

\subsection{Vital Sign Estimation}
We consider 
two sources $q\in\{r,h\}$ in the estimated phase signal ${\tilde x}_m$ of the vital bin, contributed from respiration and heartbeat, while each is composed of $L$ harmonics from the model assumption as in \eqref{eq:vs_signal}. To observe signal frequency, we construct a continuous $M$-point observation window, where $2L < M \leq N_s$, the estimated vital-sign-related phase signal is expressed as:
\begin{align}
    \tilde {\mathbf x}_m = \sum_{q\in\{r,h\}} \mathbf {Z}_q \begin{bmatrix}
        e^{j \omega_q 1 m} & & 0 \\
         & \ddots & \\
        0 &  & e^{j \omega_q L m} \\
    \end{bmatrix}
    \mathbf{m}_q + \mathbf w_m .
\end{align}
$\tilde {\mathbf x}_m = \left[ {\tilde x}_m, {\tilde x}_{m-1}, ..., {\tilde x}_{m-M+1} \right]^T \in \mathbb C^{M \times 1}$ represents the slow-time sampled signal of length $M$ in a observation window of time index $m$. 
$\mathbf {Z}_q = [\mathbf {z}(\omega_q 1), \mathbf {z}(\omega_q 2), \dots, \mathbf {z}(\omega_q L) ] \in \mathbb C^{M \times L}$ represents the observation signal composed of $L$ harmonics, where $\mathbf {z}(\omega_q) = [1, e^{j\omega_q}, e^{j2\omega_q}, e^{j(M-1)\omega_q} ]^T \in \mathbb C^{M \times 1}$. 
$\mathbf{m}_q = [{M}_{q_1}, {M}_{q_2}, \dots, {M}_{q_L}]^T  \in \mathbb C^{L \times 1}$
denotes the complex amplitude of each harmonic. $\omega_q$ denotes the fundamental frequency of the $q$-th source signal, either from respiration ($q = b$) or heartbeat ($q=h$). 
The signal is corrupted by noise $\mathbf w_m = [w_m, w_{m+1}, ..., w_{m+M-1}]^T \in \mathbb C^{M \times 1}$, where the noise variance is $\sigma^2$. 
The covariance matrix $\mathbf R = \mathbb {E}\{\tilde {\mathbf x}_m \tilde {\mathbf x}^H_m\}$ and its eigenvalue decomposition can be estimated as:
\begin{align}
    \hat {\mathbf R} = \frac{1}{N_s - M + 1} \sum_{m=0}^{N_s-M} \tilde {\mathbf x}_m \tilde {\mathbf x}_m^H = \mathbf {U} \pmb {\Lambda} \mathbf {U}^H \ ,
\end{align}
where $\mathbf U = [\mathbf u_1, \mathbf u_2, ..., \mathbf u_M]$ forms the matrix of normalized eigenvectors of this correlation matrix, and $\pmb \Lambda$ is the diagonal matrix of eigenvalues corresponding to each eigenvector. 
To distinguish between signal and noise components, let $\mathbf{G} = [\mathbf{u}_{2L+1}, \mathbf{u}_{2L+2}, ..., \mathbf{u}_{M}]$ represent the subspace composed of $M-2L$ noise eigenvectors. The fundamental frequency components of respiration and heartbeat can be estimated by the proposed harmonic MUSIC (HMUSIC) as:
\begin{align}
    \{ \hat{\omega}_q \} = \arg \max_{ \{ {\omega}_q \in [{\omega}_{q_{min}}, {\omega}_{q_{max}}] \} }  \frac {1} {\sum_{q\in\{r,h\}} \|\mathbf {Z}_q^H \mathbf {G}\|_{\text F}^2  } \ ,
\end{align}
where $[{\omega}_{q_{min}}, {\omega}_{q_{max}}]$ represents the frequency range for the desired frequency signal of the $q$-th source. 

\section{Experimental Results}

\subsection{mmWave Radar Configuration}

We utilized the Texas Instruments (TI) industrial FMCW radar module, IWR6843ISK, and the radar's mixed-signal data was sampled by an ADC using the DCA1000EVM. Using the UDP protocol, this data was transmitted in real-time to a computer via Ethernet. 
In our experimental setup, the height of the individuals' chest was aligned with the height of the radar antenna. 
Since our discussions primarily focused on horizontal angle target localization, we used 2 horizontal transmitting antennas and 4 receiving antennas, employing TDM-based MIMO \cite{TDM_MIMO} with equivalent $N_v = 8$ virtual antennas in the azimuth direction. 
The radar parameters for our experiments were configured based on the experimental scenario, as detailed in Table \ref{tab:radar_setting}.

\begin{table}
    \centering
    \caption{mmWave Radar Configurations}
    \label{tab:radar_setting}
    \begin{tabular}{||l c c||} 
     \hline
     Parameter & Value & Unit\\ [0.5ex] 
     \hline\hline
     Start Frequency $f_0$ & 60 & GHz \\ 
     \hline
     ADC Samples $N_f$ & 240 & samples \\ 
     \hline
     Chirp Duration $T_c$ & 66.66 & $\mu$s \\ 
     \hline
     Frequency Slope $S$ & 60 & MHz/$\mu$s \\ 
     \hline
     Valid Bandwidth & 3600 & MHz \\ 
     \hline
     ADC Sample Rate & $4000 \cdot 10^3$ & samples per second \\ 
     \hline
     Frame Duration $T_s$ & 50 & ms \\ 
     \hline
    \end{tabular}
\end{table}

\begin{figure}
    \centering
    \subfigure[Respiration Rate Estimation Accuracy]{
        \includegraphics[width=.8\linewidth]{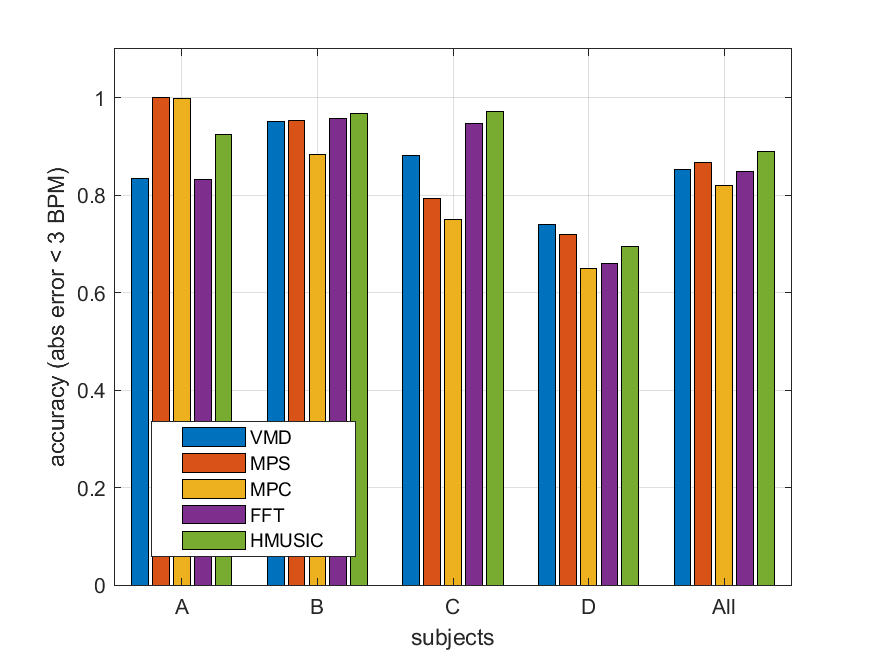}
    }
    \subfigure[Heart Rate Estimation Accuracy]{
        \includegraphics[width=.8\linewidth]{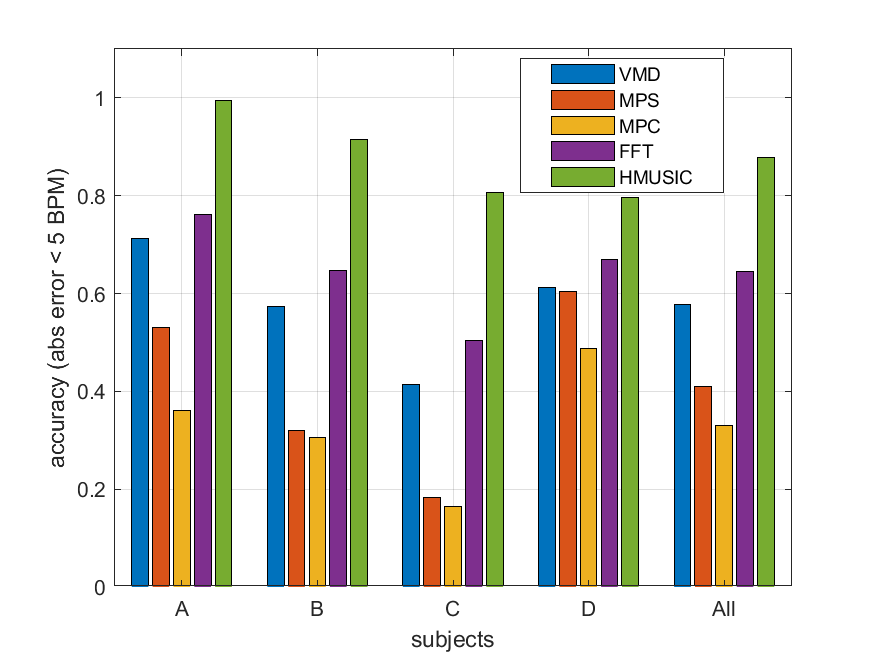}
    }
    \caption{Accuracy of Vital Signs Estimation for Each Subject}
    \label{fig:single_acc_sub}
\end{figure}

\subsection{Data Capture Setting}
To validate the results of vital sign estimation, this study employed Vernier's Go Direct respiration belt and electrocardiography (EKG) sensors as ground truth. 
These sensors measured the force exerted on the abdomen during respiration and electrocardiogram data, which were wirelessly transmitted in real-time to a computer via Bluetooth.

To ensure sufficient resolution for respiration and heartbeat frequencies and maintain a reasonably stable measurement process, general vital sign estimation was conducted within a 10 to 15 seconds window to cover a period. In this study, a total of $N_s = 256$ slow-time samples (equivalent to 12.8 seconds) were used in each estimation. Each experiment involved 2400 slow-time samples (equivalent to 2 minutes) for result assessment. 

Time synchronization between the radar and reference sensors was achieved during data collection on the computer, ensuring that the total data length difference between the two data sets was less than a 2-sample error in slow time (equivalent to 100 milliseconds). A sufficiently large estimation time window was used to disregard synchronization errors between the reference sensor and radar. If the data length difference exceeded 100 milliseconds, it indicated a possible delay or transmission failure in the Bluetooth-transmitted data, and such data were not used in the analysis.

\subsection{Vital Sign Estimation of Different Subjects} \label{sec:vs}

In Fig. \ref{fig:single_acc_sub}, we discuss the accuracy of vital sign estimation for four individuals and compare the respiration rate and heart rate estimation errors for each subject using various methods. Compared to sensors' ground value, the accuracy for respiration rate (RR) is defined as the empirical probability of breathing rate errors less than $3$ breaths per minute (BPM), and for heart rate (HR), it's defined as the probability of heart rate errors less than $5$ beats per minute (BPM). We compare our method with reference methods including VMD \cite{ref3_mmHRV}, MPS \cite{ref4_MPS}, MPC \cite{ref5_MPC}, and FFT \cite{ref8_TI}.

Since respiration signals are a primary component of vital signs with larger amplitude than heartbeat signals, there may not be significant differences in the accuracy of respiration rate among the methods. However, there are noticeable differences in the comparison of heart rate, with the proposed method achieving a heart rate accuracy of $0.88$ for all subjects. 
Since the phase signal is not sinusoidal, VMD is less efficient than other methods such as FFT. 
Besides, MPC relies on the optimal bin selection, achieving amplitude and phase coherence, but the property seems not to hold given the cluttered environment. 
Although respiration harmonic rejection is designed in MPS, it only shows a better result in subject D. We infer its lack of statistical results from the most frequent frequency in the heart rate pseudo-spectrum, which differs for different subjects to degrade the performance.

\section{Conclusion}

This paper studies vital sign estimation of individuals using mmWave MIMO FMCW radar. To extract the fundamental frequencies of respiration and heartbeat from vital sign signals, a harmonic MUSIC (HMUSIC) algorithm is proposed to overcome the problem of harmonic components of respiration interfering with heartbeat estimation and obtain a high-resolution pseudo-spectrum. Experimental observations in practical scenarios involving $4$ different subjects demonstrate that the multi-vital bin design of the proposed method offers improvements over existing techniques, particularly enhancing the accuracy of heart rate estimation in a single range scenario, with respiratory rate accuracy around $0.89$ and heart rate accuracy approximately at $0.88$. 

\bibliographystyle{IEEEbib}
\bibliography{IEEEabrv2.bib}

\end{document}